# Extension of a Pattern Recognition Validation Approach for Noisy Boson Sampling

Yang Ji*[1,2], Yongzheng Wu[1,2], Shi Wang[1,2], Jie Hou[1,2], Meiling Chen[1], Ming Ni[1]
[1]*The 32nd Research Institute of China Electronics Technology Group Corporation, Shanghai, 201808, China*
[2]*CAS Centre for Excellence and Synergetic Innovation Centre in Quantum Information and Quantum Physics, University of Science and Technology of China, Shanghai, 201315, China*
*Corresponding authors. E-mail: yangjimtz@163.com

Boson sampling is one of the main quantum computation models to demonstrate the quantum computational advantage. However, this aim may be hard to realize considering two main kinds of noises, which are photon distinguishability and photon loss. Inspired by the Bayesian validation extended to evaluate whether distinguishability is too high to demonstrate this advantage, the pattern recognition validation is extended for boson sampling, considering both distinguishability and loss. Based on clusters constructed with the K means++ method, where parameters are carefully adjusted to optimize the extended validation performances, the distribution of characteristic values is nearly monotonically changed with indistinguishability, especially when photons are close to be indistinguishable. However, this regulation may be suppressed by photon loss. The intrinsic data structure of output events is analyzed through calculating probability distributions and mean 2-norm distances of the sorted outputs. An approximation algorithm is also used to show the data structure changes with noises.

Boson sampling was proposed to give evidence that the extended Church-Turing thesis is not true [1]. It is also potential in designing drugs [2-5] and solving graph problems [6-8]. The feasibility of boson sampling validations has been in debate [9] because of the computation complexity until the proposal of the row-norm estimation [10], which gives strong evidence that boson sampling is far from uniform.

The Bayesian validation is a more universally practical approach [11-16], where the complex matrix permanent needs to be calculated once for each output event and several events are needed. The reliability of an ideal Bayesian validation for boson sampling is threatened by sampling with partial indistinguishable photons. In certain cases, samples may also pass an ideal Bayesian validation while photon indistinguishability does not meet needs of the quantum computational advantage. The Bayesian validation was further extended to evaluate whether distinguishability is too high to demonstrate this advantage [17]. In the extended approach, a stable slope threshold is formed through training events from a boson sampler where indistinguishability ($x_{\mathrm{ind}}$) exactly meets quantum computational advantage needs, that is, $x_{\mathrm{ind}}$=0.947. One only needs to obtain a slope with a boson sampler to be tested to learn whether $x_{\mathrm{ind}}$ has met the needs through comparing the slope with the threshold.

The calculation of a Haar-random matrix permanent has a computation complexity of $O(n2^n)$ using the Ryser algorithm with the Gray code [18-20], where $n$ is the input photon number. Therefore, based on the Bayesian validation, it is hard to distinguish large-scale boson sampling from other distributions. As a contrast, the pattern recognition validation [21] may be more suitable for large-scale boson sampling because of avoidances of permanent calculations if there is a bona fide sampler. Inspired by the extension of the Bayesian validation, we extend the pattern recognition validation in order to tell whether noises are too strong to demonstrate the quantum computational advantage.

The ideal boson sampling is a quantum computation model where $n$ indistinguishable photons enter an $m$-mode interferometer. The interferometer could be consisted of a series of beam splitters and phase shifters [22] and is corresponding to an $m$-dimensional Harr-random matrix $\mathbf{M}$ mathematically. The output probability is [18]

$$\Pr(T) = \frac{|\operatorname{Perm}(\mathbf{M}^{S,T})|^2}{\prod_{i=1}^{m} S_i! \prod_{i=1}^{m} T_i!}. \tag{1}$$

$|S\rangle = |S_1, S_2, \ldots, S_m\rangle$ and $|T\rangle = |T_1, T_2, \ldots, T_m\rangle$ are input and output Fock states, respectively. $\mathbf{M}^{S,T}$ is the submatrix of $\mathbf{M}$ where the $i$th row is repeatedly chosen for $S_i$ times and the $i$th column is repeatedly chosen for $T_i$ times. $\operatorname{Perm}(\mathbf{M}^{S,T})$ is the permanent of $\mathbf{M}^{S,T}$. The output combinations will become nearly collision-free if $m \propto n^2$, where $T_i$ is 0 or 1 [23].

When distinguishability is taken into consideration, eq. (1) is changed as [24]

$$\Pr(T) = \frac{\sum_{\sigma}(\prod_{j=1}^{n} a_{\sigma_j j}) \operatorname{Perm}(\mathbf{M}^{S,T} \odot \mathbf{M}_{1,\sigma}^{S,T*})}{\prod_{i=1}^{m} S_i! \prod_{i=1}^{m} T_i!}, \tag{2}$$

where $\sigma$ goes over all permutations of $[n] = \{1, 2, \ldots, n\}$ and $\sigma_j$ is the $j$th element of $\sigma$, $\mathbf{M}_{1,\sigma}^{S,T}$ is the rearrangement of $\mathbf{M}^{S,T}$ where the rows are unpermuted and the columns are permuted according to $\sigma$, $\mathbf{M}_{1,\sigma}^{S,T*}$ is the complex conjugate matrix of $\mathbf{M}_{1,\sigma}^{S,T}$, $\odot$ denotes the elementwise product. $a_{ij} = \langle \psi_i | \psi_j \rangle$ is an element of the matrix of mutual distinguishability, where $\psi_i$ is the wave function of the $i$th photon. When the hypothesis that photon indistinguishability $x_{\text{ind}}$ is fixed for different input photons is taken into consideration, $a_{ij}$ could be expressed as $a_{ij} = x_{\text{ind}} + (1 - x_{\text{ind}})\delta_{ij}$. With output probabilities, boson samples considering distinguishability could be obtained based on classical simulation methods such as Markov Chain Monte Carlo (MCMC) [25, 26].

It is believed that $x_{\text{ind}}$ should be higher than 0.947 for a 50-photon boson sampler to demonstrate the quantum computational advantage, otherwise it could be approximately simulated by an efficient classical algorithm with an error threshold of 10% [27]. The Bayesian validation approach has been extended in order to tell whether $x_{\text{ind}}$ is lower than 0.947. Through training simulated collision-free events with $x_{\text{ind}}$=0.947, a stable slope of $\ln X$ could be obtained. $\ln X$ is expressed as [17]

$$\ln X = \sum_{i=1}^{N_{\text{event}}} \ln \left(\frac{\Pr(Q_i)}{\Pr_{Q-\text{CFS}}} \cdot \frac{\Pr_{C-\text{CFS}}}{\Pr(C_i)}\right), \tag{3}$$

where $N_{\text{event}}$ is the event number, $\Pr(Q_i)$ and $\Pr(C_i)$ are probabilities of the $i$th event from the ideal boson sampler and from a classical sampler such as the uniform respectively, $\Pr_{Q-\text{CFS}}$ and $\Pr_{C-\text{CFS}}$ are probability sums of collision-free outputs from these two kinds of samplers, respectively. One can tell whether $x_{\text{ind}}$ is lower than 0.947 through comparing the slope of $\ln X$ obtained with a sampler to be tested with the threshold. Inspired by the work mentioned above, we extend another kind of validation approaches, that is, the pattern recognition validation.

The pattern recognition approach can avoid permanent calculations during the validation process if there is a bona fide sampler. Consider output events of a boson sampler to be tested with $n$ partial indistinguishable photons entering an $m$-mode interferometer, where the corresponding matrix $\mathbf{M}$ can be obtained through one-photon and two-photon experimental interferences [28] or initially constructing with beam splitters and phase shifters [29, 30]. In order to match with the intrinsic data structure of output events, clusters should be constructed with clustering methods such as Bubble, K means and K means++. The later one is considered to be a robust method because of its learning capability. The steps of the pattern recognition validation based on K means++ are introduced as follows [21].

(1) A bona fide sampler is used to produce collision-free output events.

(2) Once a set of output events are generated, they are employed to construct $k$ clusters, where $k$ is set artificially. The centroid of each cluster has the mean coordinates of all the events in the cluster. The construction begins from determining the first centroid through randomly choosing from the observed events. The distances of the left events from the nearest centroid are then evaluated considering the 2-norm distance $\mathrm{L}_2(p,q)=\sqrt{\sum_{i=1}^{m}|p_i-q_i|^2}$, where $|p\rangle$ and $|q\rangle$ are Fock states of the event and the centroid, respectively. A new centroid is then chosen from the left events with a probability proportional to $L_2^2$ for each event. Through iterating this procedure $k$ centroids are initialized. Each event is then sent into the cluster who has the nearest centroid to the event, according to the 2-norm distance. Subsequently, a new set of centroids are formed because of changes of events. Eventually $k$ clusters containing several events for each one and characterized by a centroid and a radius, will become stable through repeatedly sending events and forming centroids.

(3) Events from the sampler to be tested are sent into the clusters according to their coordinates. A $\chi^2$ test is then performed where $\chi^2$ is obtained from

$$\chi^2=\sum_{i=1}^{k}\sum_{j=1}^{2}[(N_{ij}-E_{ij})^2/E_{ij}],$$
$$E_{ij}=N_iN_j/k,$$
$$N_i=\sum_{j=1}^{2}N_{ij},$$
$$N_j=\sum_{i=1}^{k}N_{ij}. \quad (4)$$

$N_{ij}$ is the event number in the $i$th cluster belong to the $j$th sampler. $\chi^2$ will follow a very different distribution if the bona fide sampler and the sampler to be tested are compatible, when compared to the case that they are incompatible.

We take the first step to extend the pattern recognition approach to tell whether photon distinguishability of a boson sampler to be tested is too high to demonstrate the quantum computational advantage. In order to learn changes of $\chi^2$ distributions, a set of collision-free samples corresponding to different $x_{\mathrm{ind}}$ are generated with the MCMC method. For each sample with $x_{\mathrm{ind}}$ fixed, the events are sent into the clusters constructed with the K means++ method using a simulated ideal boson sampler on the same scale as the bona fide sampler. Subsequently a $\chi^2$ test is performed.

The procedure of event sending and $\chi^2$ testing is repeated for several times to generate a $\chi^2$ distribution, as shown in Figure 1(a), where 4-photon 16-mode simulated samplers are employed and calculations are simplified through repeating picking up events from a box for each time. The test values of $\chi^2$ will all follow Gaussian distributions no matter they are from a classical sample or from a quantum-like sample. However, when $x_{\mathrm{ind}}$ is increased, the Gaussian center will move to a higher value in tendency. That is to say, the pattern recognition approach can not only distinguish ideal quantum samples and classical ones, but also give a clear regulation with changing $x_{\mathrm{ind}}$.

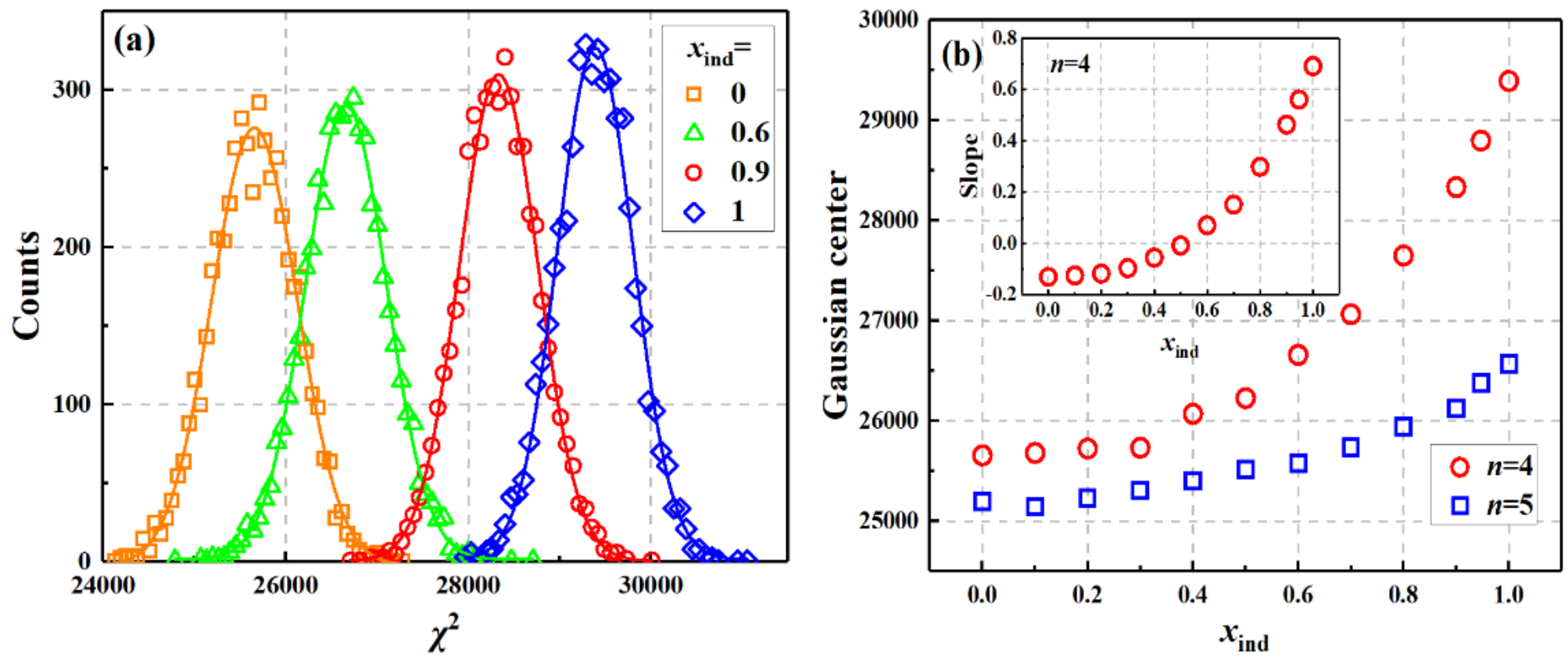

Figure 1 Validations for boson sampling considering distinguishability. (a) $\chi^2$ distributions with changing $x_{\text{ind}}$. The solid lines show the Gaussian fitting. In our scheme, clusters are initially formed using a simulated 4-photon 16-mode ideal boson sampler as the bona fide sampler, based on the K means++ method where $k$ is set as 100. The bona fide sampler shares the same $\mathbf{M}$ with other samplers and produces 1000 collision-free events. For each $x_{\text{ind}}$, the procedure of event sending and $\chi^2$ testing is performed for 5000 times. For each time, in order to reduce calculations, 1000 collision-free events are chosen randomly from a box containing $10^5$ events generated with the MCMC method. Therefore, for each $x_{\text{ind}}$, only $10^5$ events are in need to give a $\chi^2$ distribution. (b) Plots of the Gaussian center and $x_{\text{ind}}$ for samplers with different scales. The inset shows plots of the slope of $\ln X$ and $x_{\text{ind}}$ based on the extended Bayesian validation in the 4-photon cases.

As shown in Figure 1(b), the Gaussian center tends to be higher with increasing $x_{\text{ind}}$. This tendency becomes obvious when $x_{\text{ind}}$ is getting close to 1. Since the threshold of $x_{\text{ind}}$ is 0.947, it is reasonable to obtain a corresponding threshold of the center. One only needs to obtain a stable value of the corresponding Gaussian center with a boson sampler to be tested sharing the same $\mathbf{M}$ through extended pattern recognition validations. If this value is higher than the threshold, then $x_{\text{ind}}$ of the sampler to be tested should be higher than 0.947.

As a contrast, a series of 5-photon 25-mode samples are also measured in the same way, where the scale of collision-free output Hilbert space is more than 29 times to that of 4-photon 16-mode ones. The tendency of Gaussian center holds, while the amplitude of variation is shrunken. In order to enhance variations, more events may be in need to further characterize the data structure considering a larger-scale Hilbert space. Moreover, the extended Bayesian validation [17] is also performed for samples with different $x_{\text{ind}}$. In the core equation of obtaining $\ln X$ in this approach, that is, eq. (3), $\frac{\Pr_{C-\text{CFS}}}{\Pr(C_i)}$ could be simply expressed as $\binom{m}{n}$ when the collision-free uniform [31] is used as the reference sampler. The tendency of the $\ln X$ slope is also presented in the inset of Figure 1(b), which is similar to that of the extended pattern recognition validation.

In order to learn the influences of parameters with the K means++ clustering method on the extended validation, two aspects are taken into consideration, which are the value of $k$ and the kind of bona fide samplers. Both determine the cluster structure, which is important to validations. $k$ should not be too low, otherwise very different events may be sent into the same cluster, leading to a characterless structure. In the extended validation approach, $k$ is adjusted in order to optimize validation performances and give a stable threshold of the Gaussian center. When $k$ is relatively low, the centers corresponding to different $x_{\text{ind}}$ are too close when compared to the full width at half maximum (FWHM) of the Gaussian peaks, as shown in Figure 2(a). This may result in two main drawbacks for the validation performance.

(1) The center corresponding to $x_{\text{ind}}$=0.947 is so close to the neighboring center that the distance only occupies a little in the full distance from $x_{\text{ind}}$=0 to $x_{\text{ind}}$=1. It would be hard to get a good validation performance if $x_{\text{ind}}$ fluctuates slightly. On this aspect, the validation performance is evaluated by $r_1 = \frac{c_1 - c_{0.947}}{c_1 - c_0}$, where $c_1$, $c_{0.947}$ and $c_0$ are centers corresponding to $x_{\text{ind}}$=1, 0.947 and 0, respectively.

(2) The center corresponding to $x_{\text{ind}}$=0.947 is so close to the neighboring center that the distance is much shorter than FWHM of the Gaussian peaks. It would be hard to get a stable distance between two neighboring centers. On this aspect, the validation performance is evaluated by $r_2 = \frac{c_1 - c_{0.947}}{b_{0.947}}$, where $b_{0.947}$ is FWHM of the Gaussian peak corresponding to $x_{\text{ind}}$=0.947.

Higher $r_1$ and $r_2$ mean better validation performances. It is found that with increasing $k$ properly, the relative distance between Gaussian centers will become longer, as shown in Figure 2(b) and (c). Both $r_1$ and $r_2$ will increase firstly and then fluctuate slightly with $k$, as shown in Figure 2(d), where a stable validation performance is obtained when $k \geq 100$ in this case. The cluster structures with different $k$ are analyzed through counting the event number (from the bona fide sampler) in each cluster. When $k$ is relatively low, the clusters are characterless, that is, the event number of every cluster is nearly the same. Correspondingly, the curve of the cumulative event number with sorted clusters is nearly straight as shown in Figure 2(e). With increasing $k$, the clusters become characteristic, where the event number in each cluster becomes different and the cumulative event number increases more and more sharply with sorted clusters. When $k$ is high enough ($k \geq 150$), the cumulative curve is then changeless with increasing $k$. This is perhaps because, in this case, some clusters tend to break up into several minor ones which may have similar centroids. Therefore, the overall structure may be similar except that the number of clusters is different.

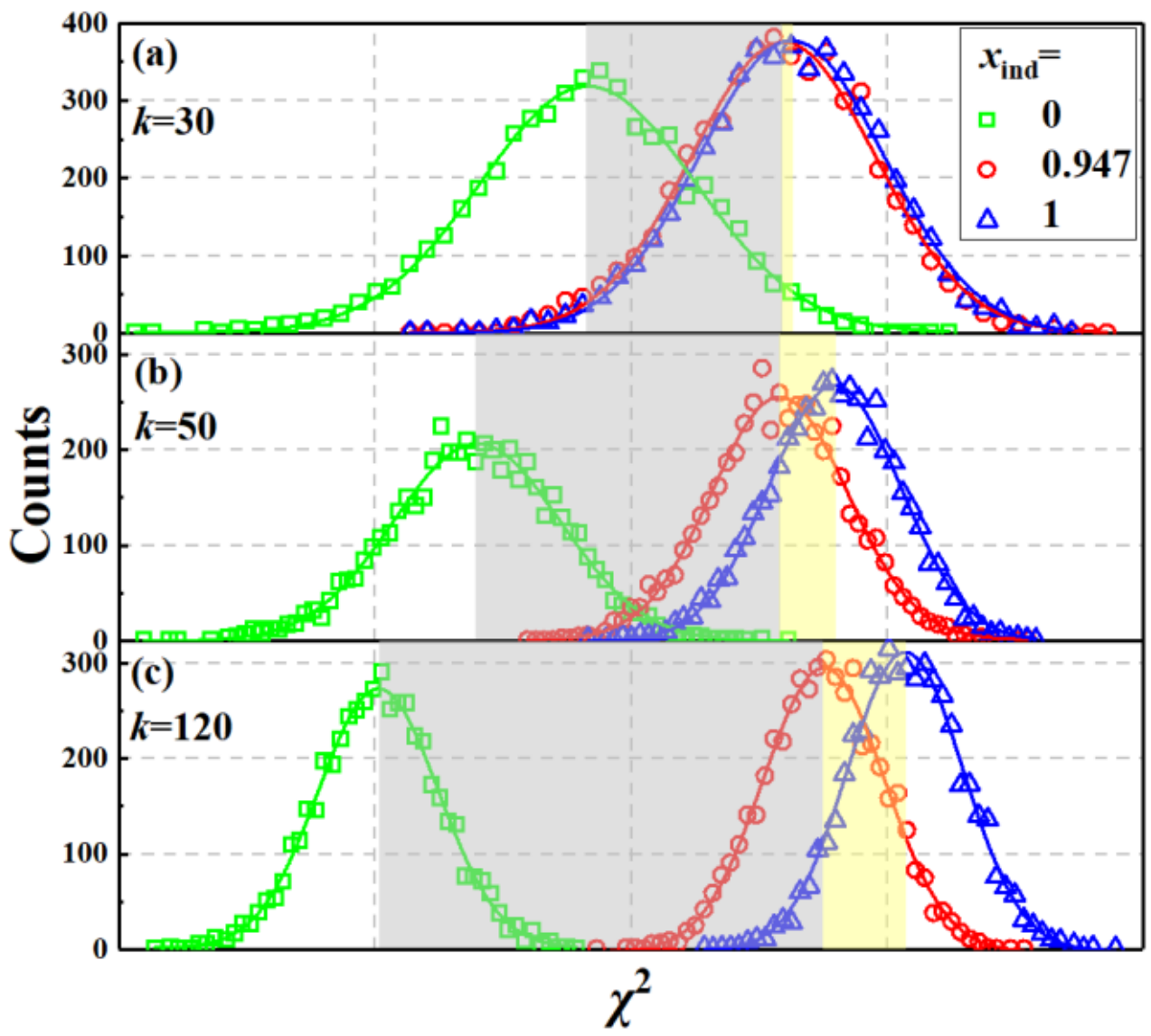

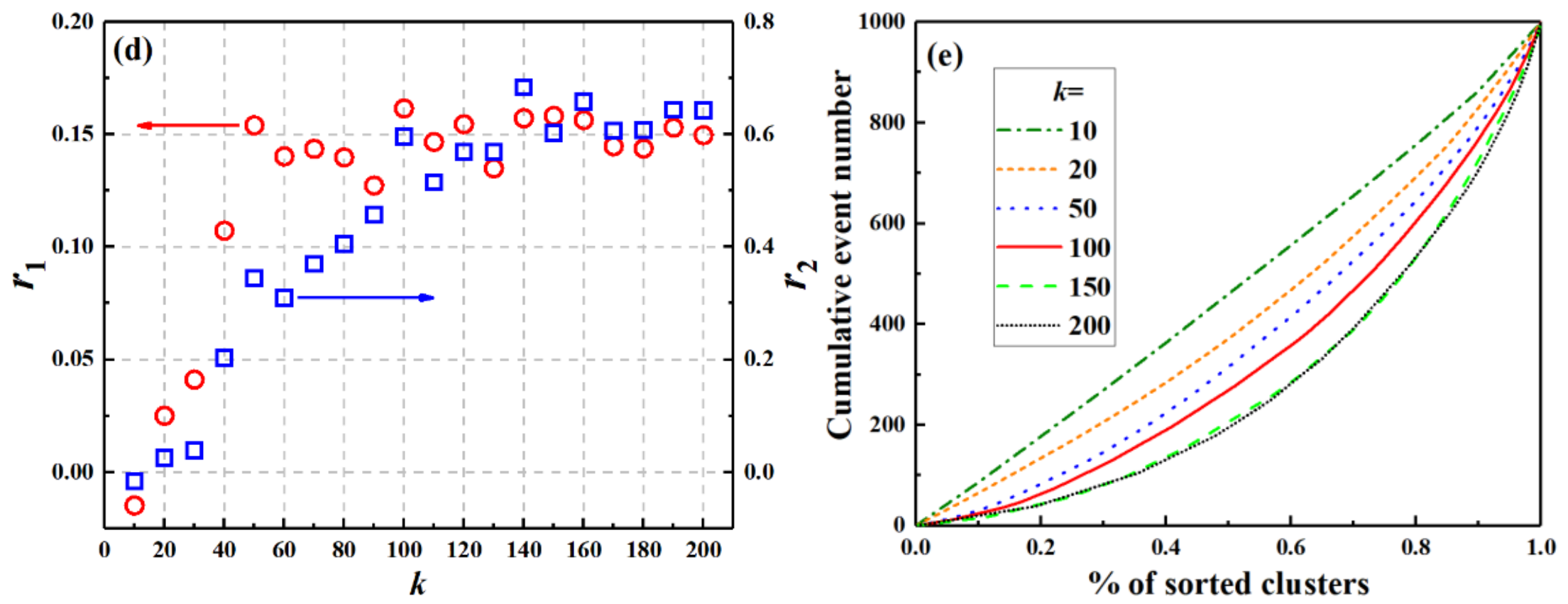

Figure 2 Validation performances with different $k$ in the 4-photon cases. (a)-(c) $\chi^2$ distributions based on clusters using the K-means++ method with $k$=30, 50 and 120, respectively. The light-grey rectangles show the distances between Gaussian centers corresponding to $x_{\mathrm{ind}}=0$ and $x_{\mathrm{ind}}=0.947$, while the pale-yellow rectangles show the distances between Gaussian centers corresponding to $x_{\mathrm{ind}}=0.947$ and $x_{\mathrm{ind}}=1$. (d) (Left) Plots of $r_1$ and $k$. (Right) Plots of $r_2$ and $k$. (e)The cumulative event number with clusters for different $k$. The clusters are sorted according to the event number in each cluster from less to more.

In the work mentioned above, a simulated ideal boson sampler is used as the bona fide sampler, contributing to a characteristic and stable cluster structure. Furthermore, a classical approximation sampler is purposely used here in order to learn the influences of bona fide sampler. In the approximation approach, the output probability is expressed as [27]

$$\Pr(T) = \frac{\sum_{j=0}^{n_{\mathrm{cutoff}}} \sum_{\sigma^j} x_{\mathrm{ind}}^j \mathrm{Perm}(\mathbf{M}^{S,T} \odot \mathbf{M}_{1,\sigma}^{S,T*})}{\prod_{i=1}^{m} S_i! \prod_{i=1}^{m} T_i!}, \tag{5}$$

where $\sigma^j$ has $j$ different elements compared to $\sigma$, and $n_{\mathrm{cutoff}}$ ($\leq n$) is the artificial cutoff photon interference number. The calculation could be simplified with keeping $n_{\mathrm{cutoff}}$ lower, although lower $n_{\mathrm{cutoff}}$ will result in a proper deviation from the ideal case. A simulated approximation sampler could be obtained using eq. (5) in combination with MCMC.

Figure 3 shows the Gaussian peaks with different $n_{\mathrm{cutoff}}$. When $n_{\mathrm{cutoff}}=n$, eq. (5) and eq. (2) actually describe the same output behavior. In this case, validations perform well with $k$ set properly, as shown in Figure 3(a). It is found that the Gaussian centers will become closer when $n_{\mathrm{cutoff}}$ is lower, as shown in Figure 3(b) and (c). This is due to the bona fide approximation sampler tends to be classical with keeping $n_{\mathrm{cutoff}}$ lower, since the higher order terms in eq. (2) present interferences of more photons and they are removed in eq. (5), despite they have a smaller weight in eq. (2) [27, 32].

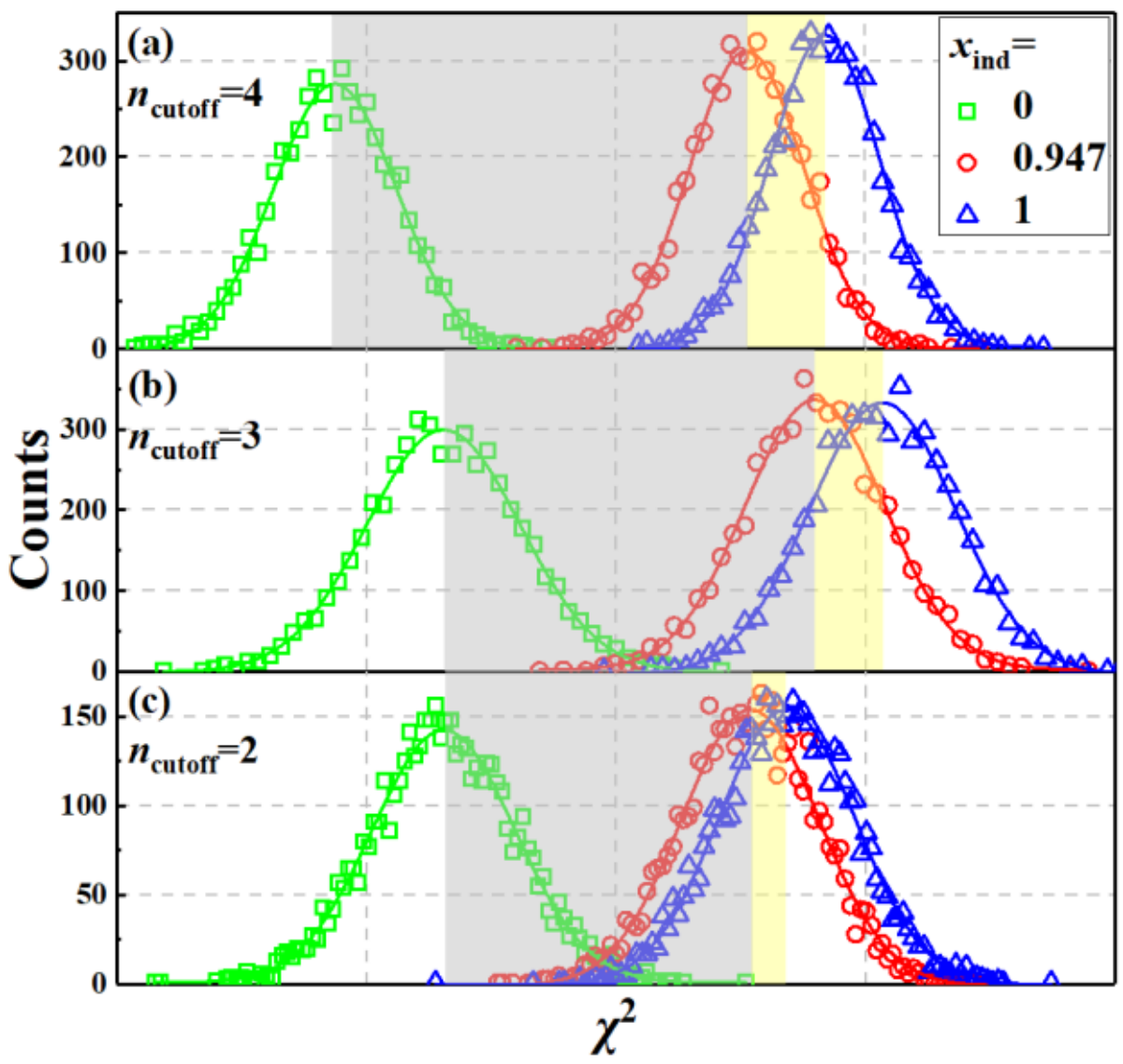

Figure 3 Validation performances with different bona fide samplers. (a)-(c) $\chi^2$ distributions are obtained based on clusters using the K-means++ method with $k$=100. An approximation of the 4-photon boson sampler is used as the bona fide sampler, keeping $x_{ind}$ as 1 and the cutoff interference photon number $n_{cutoff}$ as 4, 3 and 2, respectively.

Based on Figure 3(a)-(c), $r_1$ and $r_2$ are calculated to evaluate the validations using the bona fide approximation samplers. As presented in Table 1, with reducing $n_{cutoff}$, both $r_1$ and $r_2$ decrease accordingly, indicating the worse validation performances. When $n_{cutoff}$ is further lower, $r_1$ and $r_2$ decrease faster. It is notable that the approximation will degenerate more quickly into a classical sampler when $n_{cutoff}$ becomes further lower, since the higher order quantum interference part tends to have a smaller weight in eq. (2). Therefore, the extended pattern recognition validation can only perform well when the bona fide sampler is quantum-like.

Table 1 $r_1$ and $r_2$ with $n_{cutoff}$, in the 4-photon cases.

| $n_{cutoff}$ | $r_1$ | $r_2$ |
|---|---|---|
| 4 | 0.159 | 0.579 |
| 3 | 0.154 | 0.409 |
| 2 | 0.107 | 0.227 |

We try to explain the regular behaviors of Gaussian centers with different $x_{ind}$, through learning the intrinsic data structure of events, where output probability distributions and 2-norm distances from the event with the highest probability are considered.

Firstly, all the probabilities of $\binom{m}{n}$ collision-free outputs are calculated based on eq. (2). It is found that the output probability distribution tends to be unbalanced with increasing $x_{ind}$. When $x_{ind}$ is high, the probability of those events having a relatively high probability among all the collision-free ones will furtherly tend to be higher, especially when $x_{ind}$ is close to 1. This could be obviously observed through the cumulative probability presented in Figure 4(a).

Secondly, the mean 2-norm distance $\overline{L_2}$ is evaluated according to the sorted collision-free outputs with probabilities from high to low, where $\overline{L_2}$ is expressed as $\overline{\mathrm{L}_2}(N_{\mathrm{out}}) = \frac{\sum_{i=1}^{N_{\mathrm{out}}} \mathrm{L}_2\,(T_i,T_1)\mathrm{Pr}\,(T_i)}{\sum_{i=1}^{N_{\mathrm{out}}} \mathrm{Pr}\,(T_i)}$, where $N_{\mathrm{out}}$ is the output number, $T_i$ is the $i$th output among all the sorted outputs, $\mathrm{L}_2\,(T_i, T_1)$ is the 2-norm distance between $T_i$ and $T_1$ (the output with the highest probability), and $\mathrm{Pr}\,(T_i)$ is the probability of $T_i$. As shown in Figure 4(b), $\overline{L_2}$ is obviously different and shows the regulation with changing $x_{\mathrm{ind}}$, especially when $x_{\mathrm{ind}}$ is close to 1, indicating that the data structures of events are also different on the aspect of norm distances. In the collision-free cases, $L_2$ could only be equal to $\sqrt{2l}$, where $l$=0, 1, …, $n$. For samplers where $n$=4, we calculate the output probabilities corresponding to $L_2 \le \sqrt{2}$ and $L_2 \ge \sqrt{6}$. As shown in Figure 4(c), in both cases the probabilities increase with $x_{\mathrm{ind}}$, where the tendency becomes more obvious when $x_{\mathrm{ind}}$ is getting close to 1.

Based on the analysis mentioned above, the intrinsic data structures of output events are discussed. Despite events will normally have a higher mean probability if they have a shorter $L_2$ distance from the event with the highest probability, they only occur a little in the whole collision-free output Hilbert space. For example, for the ideal 4-photon 16-mode sampler, the mean probability of events with $L_2 \le \sqrt{2}$ is about 0.20%, which is much higher than the overall mean value of 0.05%. However, they only occur 2.69% in the whole output Hilbert space. Therefore, the overall probability will normally be higher for events with a higher $L_2$ value. When $x_{\mathrm{ind}}$ is increased, the internal (with lower $L_2$) and external (with higher $L_2$) events will correspondingly be more, while the middle events will in turn be fewer. The regular data structures may give an explanation why the extended validation approaches can work.

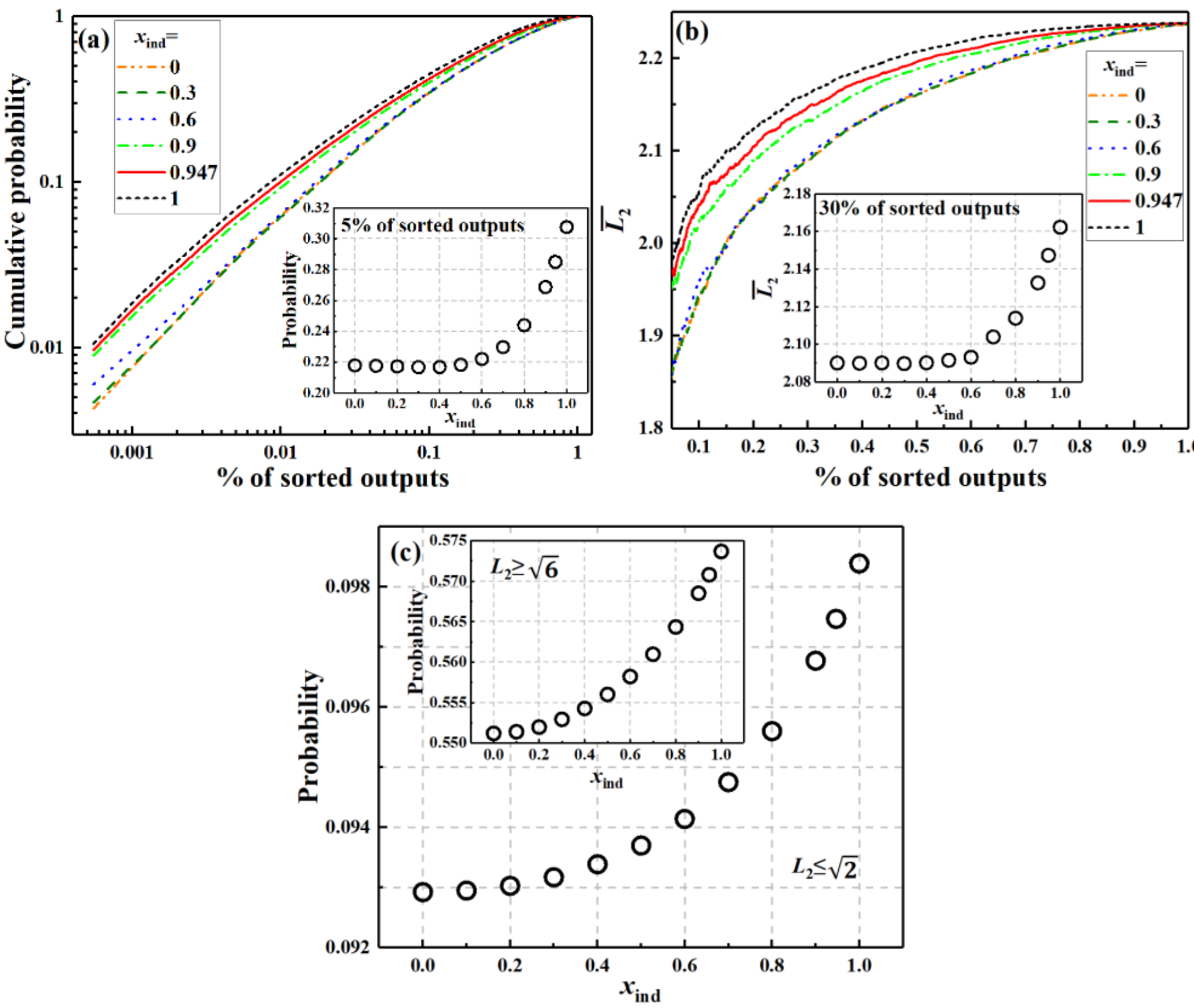

Figure 4 Intrinsic data structures of output events in the 4-photon cases. (a) Cumulative probabilities of collision-free outputs with different $x_{\mathrm{ind}}$. The outputs are sorted according to their probabilities from high to low. The inset shows overall probabilities of 5% outputs in the Hilbert space. (b) Corresponding mean 2-norm distances with $x_{\rm ind}$. The inset shows distances of 30% outputs. (c) Plots of the probability and $x_{\rm ind}$ for outputs with $L_2 \leq \sqrt{2}$ and the inset shows the cases where $L_2 \geq \sqrt{6}$.

It is worth noting that parameters such as Gaussian centers will show similar monotonical tendencies with increasing $x_{\rm ind}$, which become sharp when $x_{\rm ind}$ is close to 1. This regulation indicates that the data structure is more sensitive to higher $x_{\rm ind}$. The explanation may be given by eq. (5), which provides a comparison between the sampler with partial indistinguishable photons and the samplers where fewer photons definitely exhibit interferences. The comparison is realized through the total variation distance $D = \frac{1}{2}\sum_T \left|\mathrm{Pr}_{\mathrm{approxn}}(T) - \mathrm{Pr}(T)\right|$ [33], where $\mathrm{Pr}_{\mathrm{approxn}}(T)$ is obtained through eq. (5) with a proper $n_{\rm cutoff}$ ($<n$) value, and $\mathrm{Pr}(T)$ is obtained through eq. (2). As shown in Figure 5, when $x_{\rm ind}$ is low enough, there are no obvious differences between samplers, indicating that multiphoton interferences nearly do not happen in these cases. With increasing $x_{\rm ind}$, the differences become more obvious when the approximation sampler with relatively low $n_{\rm cutoff}$ is used. Only when $x_{\rm ind}$ is high enough, such as $x_{\rm ind}>0.5$ in Figure 5, the difference begins to become obvious when the approximation sampler with relative high $n_{\rm cutoff}$ is used. The total variation distance indicates that multiphoton interferences only play a main role in boson sampling when $x_{\rm ind}$ is high enough, otherwise the behavior of boson sampling will not be so sensitive to $x_{\rm ind}$ and will look more like to be classical.

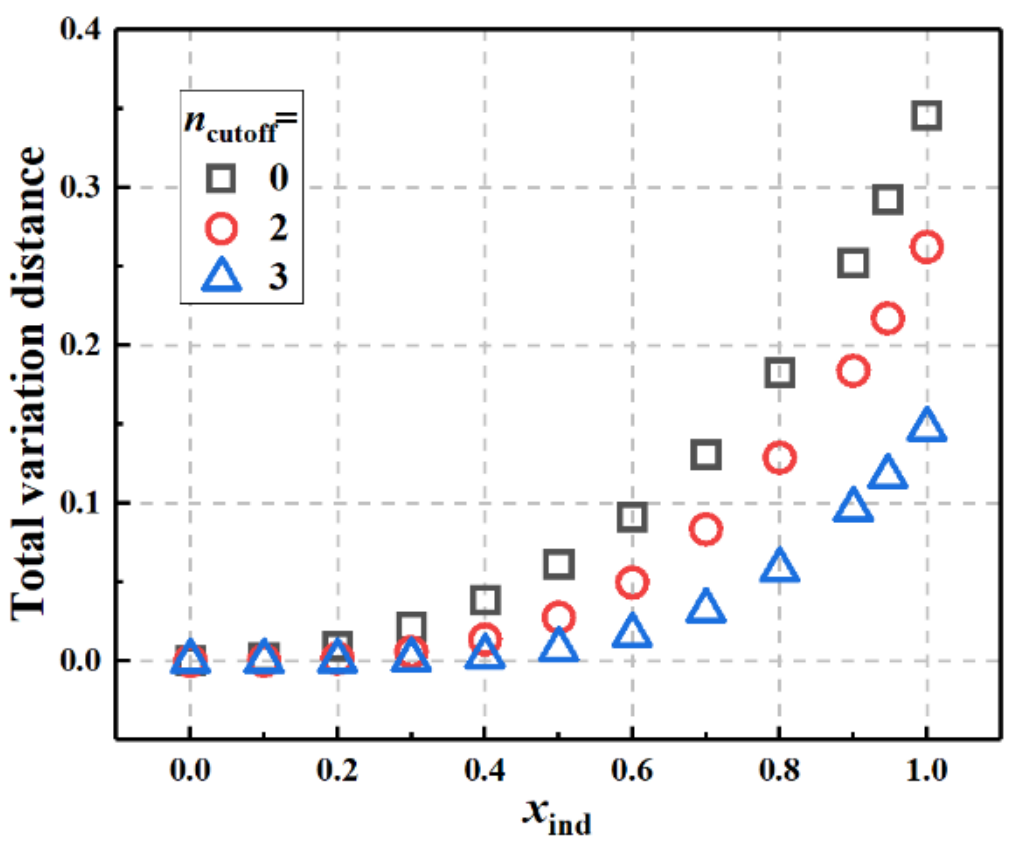

Figure 5 Total variation distances between outputs of the 4-photon boson sampler with partial indistinguishable photons and of the corresponding approximations with $n_{\rm cutoff}<n$.

In addition, another main kind of noises is also discussed, which is photon loss [11, 32, 34]. Loss is evaluated by the photon transmission $\eta = (n-l)/n$, where $l$ is the loss number, and photons may be lost in preparation, propagation and detection. It is believed that $\eta$ should be higher than 0.88 for a 50-photon boson sampler to announce the quantum computational advantage with the hypothesis that photons are indistinguishable. However, if both distinguishability and loss are considered, $\eta$ should then be replaced by $\alpha = \eta x_{\rm ind}^2$ to give an estimation, and the output probability should then simply be [32]

$$\mathrm{Pr}(T) = \binom{n}{n-l}^{-1} \sum_S \mathrm{Pr}_{\rm dis}(T|S), \tag{6}$$

where $T$ is now the output considering loss, $S$ goes over chosen input combinations having $(n-l)$ photons, resulting from the initial input state where $l$ photons are lost, and $\mathrm{Pr}_{\mathrm{dis}}(T|S)$ is the output probability of $T$ considering distinguishability based on eq. (2), with $S$ as the input.

The validation performances on 5-photon 16-mode samplers where only 4 collision-free events are detected are presented in Figure 6(a). In this case where $\eta$=0.8, the noises have already been too strong to meet the advantage needs. This case is only used here to test the extended validations. With loss, despite Gaussian centers still show a monotonical tendency, they are much closer when compared to those without loss (Figure 1). The relatively poor performances, even of the extended Bayesian validation, may be attributed to the non-determination of $S$ and the corresponding submatrix, since there are $\binom{n}{n-l}$ terms contributing to the output probability in eq. (6). This leads to data structures not so characteristic with increasing $x_{\mathrm{ind}}$. As shown in Figure 6(b) and (c), the changes of output probability distributions seem to be suppressed by loss more than those of 2-norm distances. Meanwhile, an approximation algorithm similar to that based on eq. (5) is also employed to show the sampling behaviors with both distinguishability and loss. The output probability could still be expressed by eq. (6), except that $\mathrm{Pr}_{\mathrm{dis}}(T|S)$ should now be obtained based on eq. (5) [32]. The total variation distances in Figure 6(d) show similar tendencies. However, the changes are also not so great when loss is considered. Therefore, loss will make the extended validation harder to work, where more events may be in need to give a robust performance.

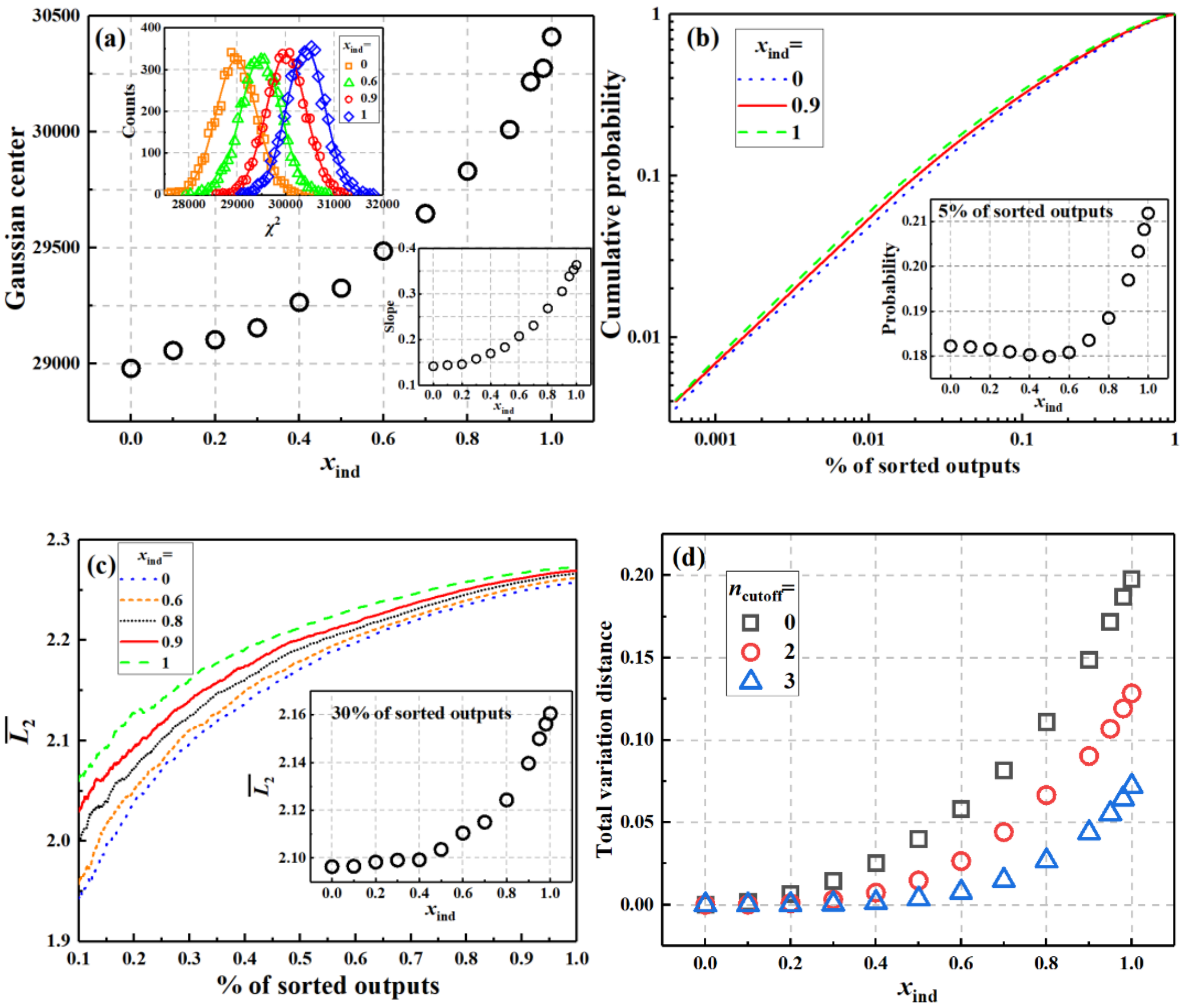

Figure 6 Cases of Boson sampling considering both distinguishability and loss. (a) Plots of the Gaussian center and $x_{\text{ind}}$, where extended validations are carried out on 5-photon 16-mode samplers with 1 photon lost. The validation parameters are kept the same with those used in Figure 1. The inset on the top left shows the Gaussian peak locations and the inset on the bottom right shows the extended validation effect based on the Bayesian approach. (b)-(c) Cumulative probabilities and mean 2-norm distances with $x_{\text{ind}}$, where the insets show the tendencies of certain values. (d) Total variation distances between outputs of the noisy boson samplers and of the approximations.

Last but not least, we give a brief outlook that the extended validations on Gaussian boson sampling, which is an important model for the easier realization of the quantum computation advantage and for practical applications [35]. The output probability is determined by the Hafnian of a submatrix containing information of the Gaussian input state, the interferometer and the output state. The Gaussian state will naturally make the submatrix more complex to construct and more unstable than that of boson sampling, with additional noises such as the phase mismatch [33]. Besides, the output photon number is not fixed, which may also result in an intrinsic data structure harder to characterize.

In summary, the pattern recognition validation is extended to tell whether conditions of photon distinguishability and loss have met needs of the quantum computational advantage for boson sampling, through evaluating the Gaussian centers based on K-means++ clusters. The performance is attributed to the regular intrinsic data structure affected by both the noises. The data structure only changes greatly when photons are close to be indistinguishable, otherwise the behavior of boson sampling will prefer to be classical.

*Acknowledgements*. This work was supported by the Science and Technology Commission of Shanghai Municipality (No. 2019SHZDZX01), Projects of the 32nd Research Institute of China Electronics Technology Group Corporation (No. DC240554-00, No. AG231031-00 and No. AG230639-00).

# References

1. Aaronson S., Arkhipov A.: The computational complexity of linear optics, in *The Computational Complexity of Linear Optics: STOC'11 Proceedings of the Forty-Third Annual ACM Symposium on Theory of Computing*, San Jose, California, USA, June 06-08, 2011 (ACM, New York, 2011), pp. 333-342.
2. Yu S., Zhong Z. P., Fang Y., Patel R. B., Li Q. P., Liu W., Li Z., Xu L., Sagona-Stophel S., Mer E., Thomas S. E., Meng Y., Li Z. P., Yang Y. Z., Wang Z. A., Guo N. J., Zhang W. H., Tranmer G. K., Dong Y., Wang Y. T., Tang J. S., Li C. F., Walmsley I. A., Guo G. C.: A universal programmable Gaussian Boson Sampler for drug discovery, Nat. Comput. Sci. 3, 839 (2023).
3. Banchi L., Fingerhuth M., Babej T., Ing C., Arrazola J. M.: Broadband quadrature-squeezed vacuum and nonclassical photon number correlations from a nanophotonic device, Sci. Adv. 6, eaax1950 (2020).
4. Wang C. S., Curtis J. C., Lester B. J., Zhang Y., Gao Y. Y., Freeze J., Batista V. S., Vaccaro P. H., Chuang I. L., Frunzio L., Jiang L., Girvin S. M., Schoelkopf R. J.: Efficient multiphoton sampling of molecular vibronic spectra on a superconducting bosonic processor, Phys. Rev. X 10, 021060 (2020).

5. Huh J., Guerreschi G. G., Peropadre B., McClean J. R., Aspuru-Guzik A.: Boson sampling for molecular vibronic spectra, Nat. Photon. 9, 615 (2015).
6. Deng Y. H., Gong S. Q., Gu Y. C., Zhang Z. J., Liu H. L., Su H., Tang H. Y., Xu J. M., Jia M. H., Chen M. C., Zhong H. S., Wang H., Yan J., Hu Y., Huang J., Zhang W. J., Li H., Jiang X., You L., Wang Z., Li L., Liu N. L., Lu C. Y., Pan J. W.: Solving graph problems using gaussian boson sampling, Phys. Rev. Lett. 130, 190601 (2023).
7. Brádler K., Dallaire-Demers P. L., Rebentrost P., Su D., Weedbrook C.: Gaussian boson sampling for perfect matchings of arbitrary graphs, Phys. Rev. A 98, 032310 (2018).
8. Arrazola J. M., Bromley T. R.: Using Gaussian boson sampling to find dense subgraphs, Phys. Rev. Lett. 121, 030503 (2018).
9. Gogolin C., Kliesch M., Aolita L., Eisert J.: Boson-sampling in the light of sample complexity, arXiv:1306.3995.
10. Aaronson S., Arkhipov A.: Bosonsampling is far from uniform, arXiv:1309.7460.
11. Wang H., Li W., Jiang X., He Y. M., Li Y. H., Ding X., Chen M. C., Qin J., Peng C. Z., Schneider C., Kamp M., Zhang W. J., Li H., You L. X., Wang Z., Dowling J. P., Höfling S., Lu C. Y., Pan J. W.: Toward scalable boson sampling with photon loss, Phys. Rev. Lett. 120, 230502 (2018).
12. Wang H., Qin J., Ding X., Chen M. C., Chen S., You X., He Y. M., Jiang X., You L., Wang Z., Schneider C., Renema J. J., Höfling S., Lu C. Y., Pan J. W.: Boson sampling with 20 input photons and a 60-mode interferometer in a $10^{14}$-dimensional Hilbert space, Phys. Rev. Lett. 123, 250503 (2019).
13. Bentivegna M., Spagnolo N., Vitelli C., Flamini F., Viggianiello N., Latmiral L., Mataloni P., Brod D. J., Galvão E. F., Crespi A., Ramponi R., Osellame R., Sciarrino F.: Experimental scattershot boson sampling, Sci. Adv. 1, e1400255 (2015).
14. Spagnolo N., Vitelli C., Bentivegna M., Brod D. J., Crespi A., Flamini F., Giacomini S., Milani G., Ramponi R., Mataloni P., Osellame R., Galvão E. F., Sciarrino F.: Experimental validation of photonic boson sampling, Nat. Photon. 8, 615 (2014).
15. Bentivegna M., Spagnolo N., Vitelli C., Brod D. J., Crespi A., Flamini F., Ramponi R., Mataloni P., Osellame R., Galvão E. F., Sciarrino F.: Bayesian approach to boson sampling validation, Int. J. Quantum Inf. 12, 1560028 (2014).
16. Deng Y. H., Gu Y. C., Liu H. L., Gong S. Q., Su H., Zhang Z. J., Tang H. Y., Jia M. H., Xu J. M., Chen M. C., Qin J., Peng L. C., Yan J., Hu Y., Huang J., Li H., Li Y., Chen Y., Jiang X., Gan L., Yang G., You L., Li L., Zhong H. S., Wang H., Liu N. L., Renema J. J., Lu C. Y., Pan J. W.: Gaussian boson sampling with pseudo-photon-number-resolving detectors and quantum computational advantage, Phys. Rev. Lett. 131, 150601 (2023).
17. Dai Z., Liu Y., Xu P., Xu W. X., Yang X. J., Wu J. J.: A Bayesian validation approach to practical boson sampling, Sci. China Phys. Mech. Astron. 63, 250311 (2020).
18. Wu J., Liu Y., Zhang B., Jin X., Wang B., Wang H., Yang X.: A benchmark test of boson sampling on Tianhe-2 supercomputer, Natl. Sci. Rev. 5, 715 (2018).
19. Clifford P., Clifford R.: Faster classical boson sampling, arXiv:2005.04214.
20. Nikolopoulos G. M., Brougham T.: Decision and function problems based on boson sampling, Phys. Rev. A 94, 012315 (2016).
21. Agresti I., Viggianiello N., Flamini F., Spagnolo N., Crespi A., Osellame R., Wiebe N., Sciarrino F.: Pattern recognition techniques for boson sampling validation, Phys. Rev. X 9, 011013

(2019).
22. Crespi A., Osellame R., Ramponi R., Brod D. J., Galvão E. F., Spagnolo N., Vitelli C., Maiorino E., Mataloni P., Sciarrino F.: Integrated multimode interferometers with arbitrary designs for photonic boson sampling, Nat. Photon. 7, 545 (2013).
23. Arkhipov A., Kuperberg G.: The bosonic birthday paradox, Geometry and Topology Monographs 18, 1 (2012).
24. Tichy M. C: Interference of identical particles from entanglement to boson-sampling, J. Phys. B: At. Mol. Opt. Phys. 47, 103001 (2014).
25. Neville A., Sparrow C., Clifford R., Johnston E., Birchall P. M., Montanaro A., Laing A.: Classical boson sampling algorithms with superior performance to near-term experiments, Nat. Phys. 13, 1153 (2017).
26. Liu Y., Xiong M., Wu C., Wang D., Liu Y., Ding J., Huang A., Fu X., Qiang X., Xu P., Deng M., Yang X., Wu J.: Sample caching Markov chain Monte Carlo approach to boson sampling simulation, New J. Phys. 22, 033022 (2020).
27. Renema J. J., Menssen A., Clements W. R., Triginer G., Kolthammer W. S., Walmsley I. A.: Efficient classical algorithm for boson sampling with partially distinguishable photons, Phys. Rev. Lett. 120, 220502 (2018).
28. Spring J. B., Metcalf B. J., Humphreys P. C., Kolthammer W. S., Jin X. M., Barbieri M., Datta A., Thomas-Peter N., Langford N. K., Kundys D., Gates J. C., Smith B. J., Smith P. G. R., Walmsley I. A.: Boson sampling on a photonic chip, Science 339, 798 (2012).
29. Reck M., Zeilinger A., Bernstein H. J., Bertani P.: Experimental realization of any discrete unitary operator, Phys. Rev. Lett. 73, 58 (1994).
30. Clements W. R., Humphreys P. C., Metcalf B. J., Kolthammer W. S., Walmsley I. A.: Optimal design for universal multiport interferometers, Optica 3, 1460 (2016).
31. Brod D. J., Galvão E. F., Crespi A., Osellame R., Spagnolo N., Sciarrino F.: Photonic implementation of boson sampling: a review, Adv. Photon. 1, 034001 (2019).
32. Renema J. J., Shchesnovich V., Garcia-Patron R.: Classical simulability of noisy boson sampling, arXiv:1809.01953.
33. Zhong H. S., Peng L. C., Li Y., Hu Y., Li W., Qin J., Wu D., Zhang W., Li H., Zhang L., Wang Z., You L., Jiang X., Li L., Liu N. L., Dowling J. P., Lu C. Y., Pan J. W.: Experimental gaussian boson sampling, Sci. Bull. 64, 511 (2019).
34. Oszmaniec M., Brod D. J.: Classical simulation of photonic linear optics with lost particles, New J. Phys. 20, 092002 (2018).
35. Hamilton C. S., Kruse R., Sansoni L., Barkhofen S., Silberhorn C., Jex I.: Gaussian boson sampling, Phys. Rev. Lett. 119, 170501 (2017).